\documentclass [aps,prb,10pt,twocolumn,showpacs,superscriptaddress]{revtex4-1}

\usepackage[T2A,T1]{fontenc}
\usepackage[utf8]{inputenc}
\usepackage[english]{babel}
\usepackage{graphicx}
\usepackage{amsmath,amssymb}
\usepackage{lmodern}
\usepackage{natbib}
\usepackage{units}
\usepackage{hyperref}
\usepackage{color,soul}
\usepackage{url}
\hypersetup{colorlinks=true,citecolor={blue},linkcolor={blue},urlcolor={blue}}
\usepackage{balance}
\usepackage{textcomp}
\renewcommand{\dh}{\fontencoding{T1}\selectfont{\symbol{240}}}

\begin{document}
\title{A driven-dissipative spin chain model based on exciton-polariton condensates}

\date{\today}

\author{H. Sigur\dh sson}
\email[correspondence address:~]{helg@hi.is}
\affiliation{Science Institute, University of Iceland, Dunhagi-3, IS-107 Reykjavík, Iceland}
\author{A. J. Ramsay}
\affiliation{Hitachi Cambridge Laboratory, Hitachi Europe Ltd., Cambridge CB3 0HE, UK}
\author{H. Ohadi}
\affiliation{Department of Physics, Cavendish Laboratory, University of Cambridge, Cambridge CB3 0HE, UK}
\author{Y. G. Rubo}
\affiliation{Instituto de Energ\'{\i}as Renovables, Universidad Nacional Aut\'onoma de M\'exico, Temixco, Morelos 62580, Mexico}
\affiliation{Center for Theoretical Physics of Complex Systems, Institute for Basic Science, Daejeon 34051, Republic of Korea}
\author{T.C.H. Liew}
\affiliation{Division of Physics and Applied Physics, School of Physical and Mathematical Sciences, Nanyang Technological University 637371, Singapore}
\author{J. J. Baumberg}
\affiliation{Department of Physics, Cavendish Laboratory, University of Cambridge, Cambridge CB3 0HE, UK}
\author{I. A. Shelykh}
\affiliation{Science Institute, University of Iceland, Dunhagi-3, IS-107 Reykjavík, Iceland}
\affiliation{ITMO University, St.\ Petersburg 197101, Russia}

\begin{abstract}
An infinite chain of driven-dissipative condensate spins with uniform nearest-neighbor coherent coupling is solved analytically and investigated numerically. Above a critical occupation threshold the condensates undergo spontaneous spin bifurcation (becoming magnetized) forming a binary chain of spin-up or spin-down states. Minimization of the bifurcation threshold determines the magnetic order as a function of the coupling strength. This allows control of multiple magnetic orders via adiabatic (slow ramping of) pumping. In addition to ferromagnetic and anti-ferromagnetic ordered states we show the formation of a paired-spin ordered state $\left|\dots \uparrow \uparrow \downarrow \downarrow \dots \right. \rangle$ as a consequence of the phase degree of freedom between condensates.
\end{abstract}

\maketitle

Many-body spin systems, both classical and quantum, have found applications in a number of fields of rising complexity. Their Hamiltonians (Ising, $XY$, Heisenberg, Sherrington-Kirkpatrick, etc.) have been used to study collective behaviors such as familiarity recognition in neural networks~\cite{Hopfield625}, hysteresis in DNA interactions~\cite{Vtyurina03052016}, combinatorial optimization problems in logistics, patterning, and economics~\cite{stein_spin_2013, Bouchaud2013}. Besides their wide application, controllable spin lattices also offer insight into physical problems such as frustration~\cite{Kim_FrustIsingNature_2010, diep2013frustrated}, spin-ice~\cite{Nisoli_SpinIceReview_2013, Drisko_spinice_2017}, spin-wave dynamics~\cite{Katine_SpinWavePillars_2000, Glazov_SpinWaveMC_2015}, domain wall motion~\cite{Stenger_SpinDomains_1998, Liew_PolNeurons_2008, Adrados_PolBullets_2011}, and spin-glass formation~\cite{ Spin_Glass_Theory_1986, stein_spin_2013}. A driven-dissipative spin lattice, where both phase and spin of the vertices are free, has yet to be addressed. Here, in contrast to entropy and minimum energy principles (as in the Ising model), the stationary physics of the system is governed by the balance of gain and decay with remarkably different solutions~\cite{Ciuti_BoseHubbard_PRL2013}. Currently, only limited investigation has been devoted to driven-dissipative lattice systems where recent works have proposed ``simulators'' based on interacting exciton-polariton condensates~\cite{Berloff_Arxiv_PolaritonSimulators} and Ising machines with degenerate optical parametric oscillators~\cite{Inagaki2016}.

Nonresonantly excited spinor exciton-polariton (or simply {\it polariton}) condensates~\cite{Fraser2016, Sanvitto2016} have developed into a popular platform for cutting edge opto-electronic and opto-spintronic technologies~\cite{Liew_polaritonDevices_2011, Sanvitto_polaritonDevices_2016}. The driven-dissipative condensates are realized by matching the gain and the decay of polaritons through continuous external driving of either optical or electrical nature. These macroscopic coherent states possess a spin and a phase degree of freedom, strong nonlinearities and a small effective mass, allowing them to interact and synchronize with other spatially separate condensates over long distances (hundreds of microns)~\cite{Snoke_PolaritonFlow_PRX2013}, making them interesting candidates for driven-dissipative spin lattices. Recently it was reported that a spinor polariton condensate bifurcates at a critical pump intensity into either of two highly circularly polarized states using a continuous linearly polarized nonresonant excitation~\cite{ohadi_spontaneous_2015}. The emission polarization is explicitly related to the polariton condensate pseudospin orientation (from here on {\it spin})~\cite{kavokin_pseudospin}. The system has since then been extended to polariton condensate spin pairs~\cite{ohadi_tunable_2016} which can controllably display alignment of antiferromagnetic (AFM) and ferromagnetic (FM) nature. This spin degree of freedom offers a unique way to study ordering amongst coupled spin vertices in various lattices.
\begin{figure}[!t]
  \centering
		\includegraphics[width=0.45\textwidth]{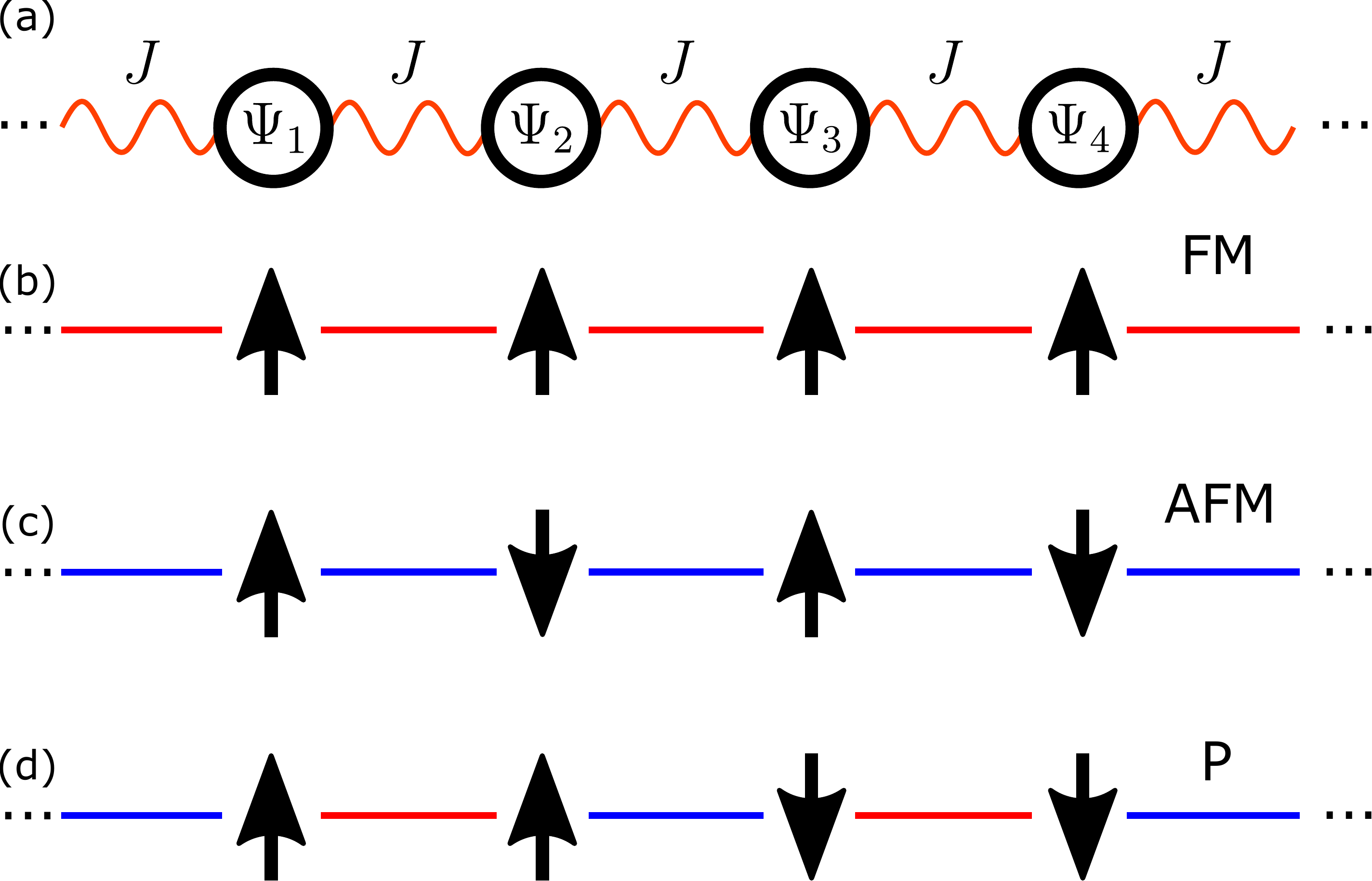}
	\caption{(Color online) (a) Schematic showing spinor condensates coupled together through the same-spin coupling parameter $J$ in an infinite chain. States with equal number of bond types per condensate can be categorized as (b) FM, (c) AFM, and (d) paired (P).}
\label{fig.scheme}
\end{figure}

In this paper, we extend such polariton condensates to an infinite chain model and present methods of controllably producing different spin-ordered chains. We solve exactly and numerically analyze the stationary states of the infinite chain of spin-bifurcated condensates with nearest-neighbor same-spin coupling in the tight binding approach. The stationary solutions correspond to ferromagnetic, antiferromagnetic, and paired-spin order states of two-up and two-down spins (P) (see Fig.~\ref{fig.scheme}). States characterized by FM bonds with zero phase-slip and AFM bonds with $\pi$ phase-slip are shown to have a minimum bifurcation threshold, and are stable against long-wavelength fluctuations. Monte-Carlo trials with adiabatic ramping of the pump intensity on a cyclic system of 4 condensates give a phase diagram in full agreement with the predicted minimum threshold winners as a function of coupling strength. This clear hierarchy for the probability of formation is an important prerequisite for a spin-lattice simulator. Non-adiabatic trials on the other hand result in a complex phase diagram, as a result of the initial condition progressing to its nearest phase space attractor. In addition to spatially uniform stationary states, we find that frustrated or defect states, with oscillating spinors can appear in this system.

Condensation of the bosonic spin $\pm1$ quasiparticles known as exciton-polaritons~\cite{kasprzak_bose-einstein_2006, balili_bose-einstein_2007, lai_coherent_2007, carusotto_quantum_2013, Byrnes2014} is regarded as the solid-state analog of cold-atom Bose-Einstein condensates~\cite{bloch_many-body_2008}.  Its spin structure and strong interactions allow one to realize spinor condensates where macroscopic coherence and superfluid character give birth to many intriguing phenomena such as the nonlinear optical spin Hall effect~\cite{kammann_nonlinear_2012, PhysRevB.91.075305}, the formation of polariton half-vortices~\cite{Lagoudakis974}, and spontaneous symmetry breaking~\cite{PhysRevLett.109.016404}. They offer a new path towards spin manipulation~\cite{Zutic_spintronicReview_2004} with already promising results on spin-switches~\cite{AmoA_spinSwitch_2010, dreismann_sub-femtojoule_2016}, and transistors~\cite{Ballarini2013}.

Driven-dissipative polariton condensates can be accurately modeled using a coherent macroscopic spinor order parameter $\Psi = (\Psi_+, \Psi_-)^\text{T}$ where $\Psi_\pm$ are the spin-up and spin-down components respectively. Similarly to the pair of polariton condensates, where the transport of polaritons from one condensate to another can be regarded as a form of coherent coupling in the tight-binding approximation~\cite{ohadi_tunable_2016}, a system of many condensates labeled by index $n$ can be described by coupled dynamical equations
\begin{align} \notag
        i \dot{\Psi}_n & =-\frac{i}{2}
        g(S_n)\Psi_n-\frac{i}{2}(\gamma-i\epsilon)\hat{\sigma}_x\Psi_n \\  \label{eq.fullGP}
        & +\frac{1}{2}(\bar{\alpha}S_n+\alpha S_{zn} \hat{\sigma}_z)\Psi_n -  \frac{J}{2} \sum_{\langle nm \rangle} \Psi_m,
\end{align}
\begin{equation} \label{eq.Sn}
S_n \equiv \frac{|\Psi_{n+}|^2 + |\Psi_{n-}|^2}{2},
\end{equation}
\begin{equation} \label{eq.Szn}
S_{zn} \equiv \frac{|\Psi_{n+}|^2 - |\Psi_{n-}|^2}{2},
\end{equation}
where the sum is over nearest neighbors. Here we define $g(S_n) \equiv -W + \Gamma + \eta S_n$ as the pumping-dissipation imbalance, $\Gamma$ is the (average) dissipation rate, $W$ is the incoherent in-scattering (or pump rate), and $\eta$ defines the gain-saturation nonlinearity~\cite{keeling_spontaneous_2008}. The birefringence of the system corresponds to the splitting of the $XY$-polarized states  in both energy ($\epsilon$) and decay-rate ($\gamma$). The interaction parameters are written as $\bar{\alpha} = \alpha_1 + \alpha_2$, $\alpha = \alpha_1 - \alpha_2$, where $\alpha_1$ and $\alpha_2$ are the same-spin and opposite-spin polariton-polariton interaction constants, respectively. Finally, $\Re{(J)}>0$ depicts the strength of the coherent same-spin Josephson type coupling between the condensates, whereas $\Im{(J)}$ gives the dissipative coupling between the condensates~\cite{aleiner_radiative_2012}. In particular, $\Im{(J)}<0$ defines the inter-site damping to account for energy relaxation. Eq.~\eqref{eq.Sn} is the average condensate population and Eq.~\eqref{eq.Szn} is the circular polarization intensity (considered as a spin here).

The critical pump intensity for condensation in a single condensate is determined by the lowest decay rate mode and can be written,  $W_\text{lin} = \Gamma - \gamma$, resulting in a linearly polarized emission. At higher pump intensities the order parameter bifurcates into either a spin-up or spin-down state due to instability in the linearly polarized modes due to their splitting ($\epsilon+i \gamma$) and polariton-polariton interactions. For a single condensate, the critical bifurcation threshold is~\cite{ohadi_spontaneous_2015}
\begin{equation} \label{eq.Wbif}
W_\text{bif} =  W_\text{lin} +   \eta \frac{\epsilon^2 + \gamma^2}{\alpha \epsilon}.
\end{equation}
In the following, we work above this critical pump threshold such that each condensate is either in a `spin-up' or a `spin-down' state.

Formally, for identical lattice sites, the symmetry-conserving stationary solutions can be found by using the following spinor ansatz:
\begin{align} \label{eq.FM}
\Psi_{n+1}=e^{i\varphi_{n+1}}\Psi_n, \qquad \text{(FM bonds)},
\end{align}
\begin{align} \label{eq.AFM}
\Psi_{n+1}=e^{i\varphi_{n+1}}\hat{\sigma}_x\Psi_n,  \qquad \text{(AFM bonds)},
\end{align}
where ($\varphi_{n+1}$) is the phase shift moving from the condensate in question to its nearest neighbor $n+1$, and is to be determined. Eq. \ref{eq.fullGP} can now be written as:
\begin{align} \notag
    i\dot{\Psi}_n & = -\frac{i}{2}(g+i\omega_J)\Psi_n -\frac{i}{2}(\gamma - i\epsilon_{J})\hat{\sigma}_x\Psi_n \\     \label{eq:dotpsi2}
    &  +\frac{1}{2}(\bar{\alpha}S_n+\alpha S_{zn}\hat{\sigma}_z)\Psi_n.
\end{align}
This corresponds to a single condensate with complex renormalized splitting $\epsilon_J$ and energy shift $\omega_J$, arising from AFM bonds and FM bonds, respectively. The strength of these parameters depends on the relative phases between nearest-neighbors in the system. The modified parameters of a chain with two nearest neighbors can be written as
\begin{align} \label{eq.epsJ}
\epsilon_J=\epsilon+J&(\delta_{i} e^{i\varphi_{i}}+ \delta_{j} e^{i\varphi_{j}}),\\
\omega_J = -J(&(1-\delta_{i})e^{i\varphi_{i}}+(1-\delta_{j})e^{i\varphi_{j}}), \label{eq.omJ}
\end{align}
where $\delta_{i} = 1,0$ for AFM or FM bonding respectively.
\begin{figure}[!t]
  \centering
		\includegraphics[width=0.45\textwidth]{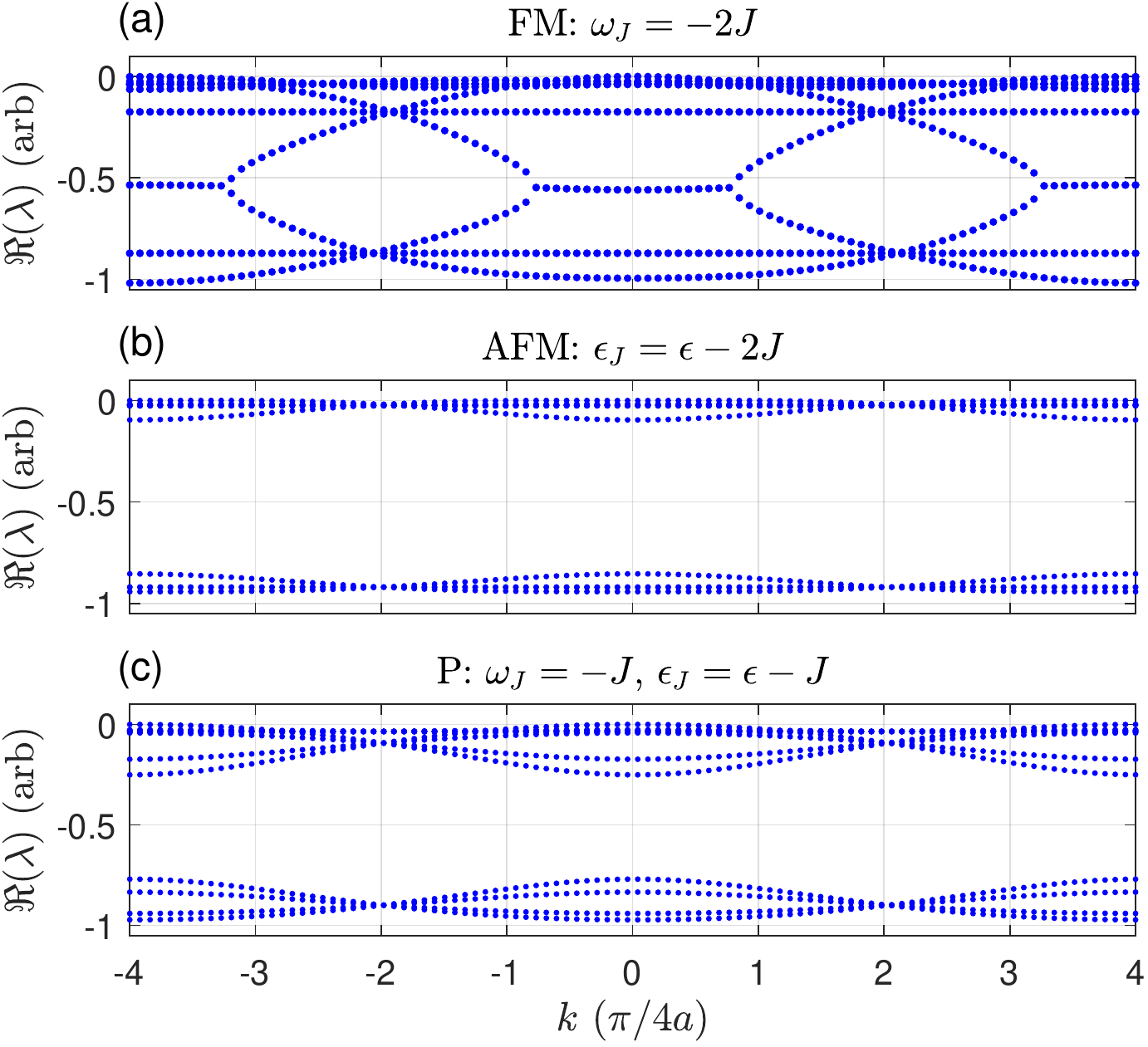}
	\caption{(Color online) Stability analysis. Plot of Lyapunov exponents $\lambda$ vs $k$-vector of fluctuations for 3 lowest bifurcation threshold chain solutions (a-c) at $W = 1.2 W_\text{bif}$, $J/\epsilon = 0.3$. For all $k$ the exponents stay negative corresponding to a stable solution.}
\label{fig.lwstab}
\end{figure}

The problem of FM- and AFM bonded condensates has thus been reduced  to a single condensate with a known solution~\cite{ohadi_spontaneous_2015}. The requirement for site-independent $\epsilon_J$ and $\omega_J$, and cyclic boundary condition of integer $2\pi$ for the phase accumulated around the closed chain, restricts the possible phases of the bonds. For chain systems of either FM or AFM ordering it can be shown that the coupling results in (see Sec.~\ref{sec.app4}):
\begin{align} \label{eq.epsomAFM}
\epsilon_J^\text{AFM} =\epsilon + 2J \cos{(2 \pi m/N)}, \qquad \omega_J^\text{AFM}=0, \\ \label{eq.epsomFM}
\epsilon_J^\text{FM}=0, \qquad \omega_J^\text{FM} = -2J \cos{(2 \pi m/N)}.
\end{align}
where $m = 0,1,2,\dots$ and $N$ is the number of condensates in the chain. For AFM chains, $N$ must be an even number since the spin unit-cell is $\left| \uparrow \downarrow \right. \rangle$.
In addition, there is a paired-spin (P) state, where each site has one FM and one AFM bond, and the spin unit-cell is $\left| \uparrow \uparrow \downarrow  \downarrow \right. \rangle$.  P solutions have $\omega_J^\text{P} = \pm J$ and $\epsilon_J^\text{P} = \epsilon \pm J$, where the signs are independent. Due to the periodicity of the stationary solutions, the essential physics of the spin-bifurcated condensate chain system can be captured within a chain of 4 condensates characterized by 10 distinct solutions (see Sec.~\ref{sec.app0}). Furthermore, we confirm the analogy between the solutions of the tight binding model (Eq.~\ref{eq.fullGP}) to a 4 condensate chain accounting for the $(x,y)$ spatial degrees of freedom (see Sec.~\ref{sec:app6}).

To identify the stable stationary solutions, we perform a long wavelength stability analysis (see Sec.~\ref{sec.app1}) for the set of coupled equations describing linear fluctuations along the periodic 4 condensate chain. Three lowest bifurcation threshold solutions of FM, AFM, and P spin order are found to have negative real-part Lyapunov exponents $\lambda$ within the first Brillouin zone of the chain (see Fig. \ref{fig.lwstab}), and hence are completely stable against fluctuations travelling along the chain. These solutions are characterized by a 0 phase slip between FM bonded condensates and $\pi$ phase slip between AFM bonded condensates (see Sec.~\ref{sec.app3} for parameter values).
\begin{figure}[!t]
  \centering
		\includegraphics[width=0.5\textwidth]{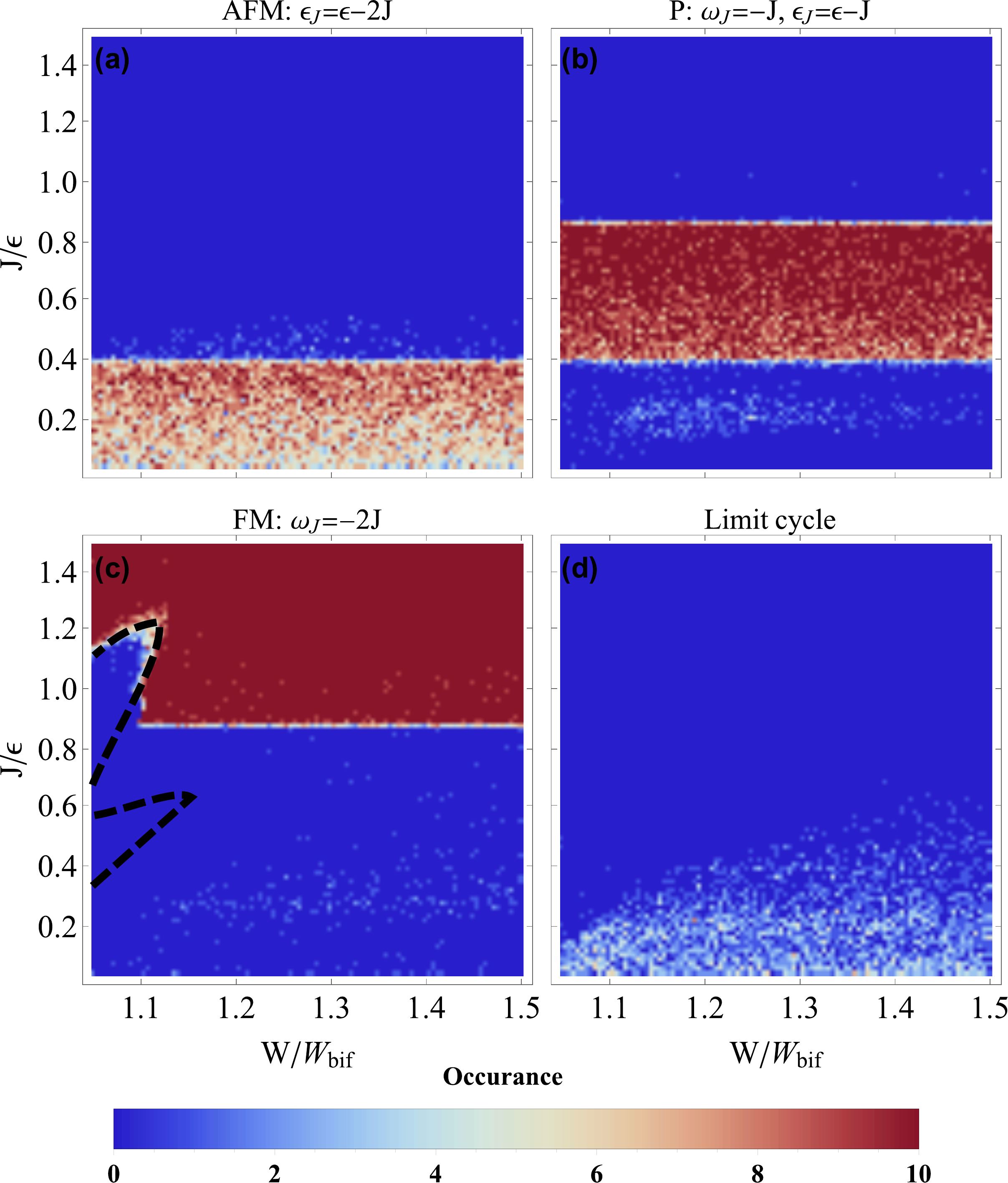}
	\caption{(Color online) Phase diagram of spin order. Probability of a spin state appearing in 10 realizations of numerical experiment, where the pump is slowly ramped to a final value $W$. (a-c) AFM, P, and FM solutions with the lowest threshold make up 87\% of the data. Black dashed lines in panel (c) indicate regimes where the solution becomes unstable. (d) A population of oscillating limit cycle solutions is noticeable (3.5\%) for low coupling strengths.}
\label{fig.raw_prob}
\end{figure}

To calculate the phase diagram and verify the analysis, we perform Monte-Carlo simulations of the periodic 4 condensate chain as a function of pump intensity $W$ and coupling strength $J$. Fig.~\ref{fig.raw_prob} shows the result of 10 Monte-Carlo (MC) trials at each site over a $100 \times 100$ pixel map in parameter space. For each realization of the numerical experiment, the  pump intensity is linearly increased from $W_0 = 0.7 W_\text{bif}$ to $W$ at a rate $10^{-4} \times  W_\text{bif}$ ps$^{-1}$, similar to the ramp-times achieved in experiments \cite{ohadi_spontaneous_2015}. Three distinct phases corresponding to the stable AFM, P, and FM stationary spin patterns are observed, as identified by the long-wavelength stability analysis.

\begin{figure}[!t]
  \centering
		\includegraphics[width=0.5\textwidth]{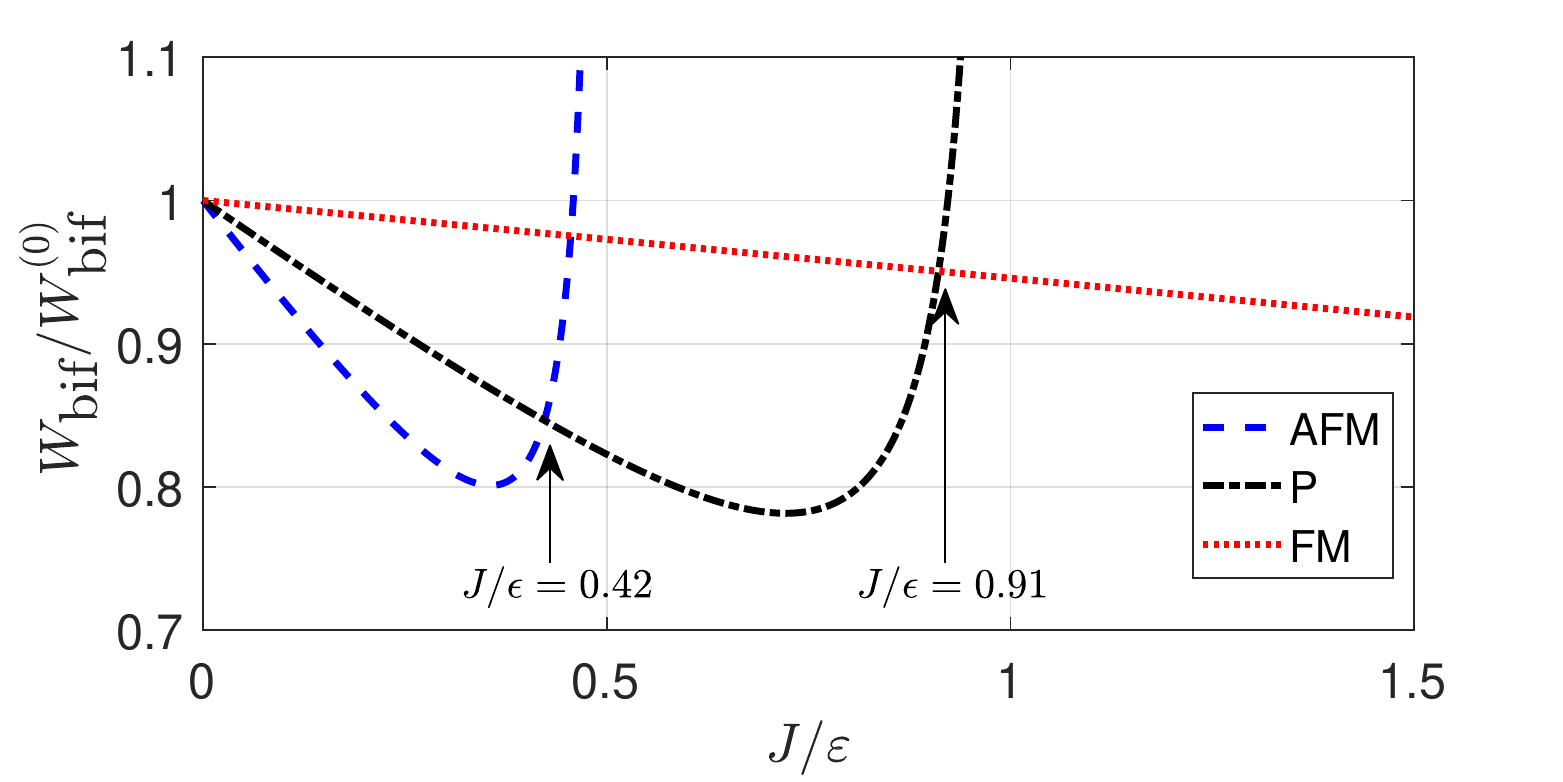}
	\caption{(Color online) Spin-bifurcation threshold vs coupling strength $J$. $W_\text{bif}^{(0)}$ is the spin-bifurcation threshold of an uncoupled condensate. The arrows indicate points where the AFM solution changes to P ($J/\epsilon = 0.42$) and P changes to FM ($J/\epsilon = 0.91$). The points are in good agreement with phase boundaries in Fig.~\ref{fig.raw_prob}.}
\label{fig.Lbif}
\end{figure}

To explain the regimes of each state (red areas in Fig.~\ref{fig.raw_prob}), we plot the spin bifurcation threshold power $W_\text{bif}$ against $J$ in Fig.~\ref{fig.Lbif}. The thresholds are calculated using Eq. \ref{eq.Wbif}, with $\epsilon\rightarrow \Re{(\epsilon_J)}$, $\gamma\rightarrow\gamma+\Im{(\epsilon_J)}$, $\Gamma\rightarrow\Gamma+\Im{(\omega_J)}$.  As the pump power is slowly increased, the state that reaches the bifurcation threshold first wins, since it has time to stabilize before competing states can bifurcate. The calculated phase-boundaries of $J/\epsilon=0.42, 0.91$ for the AFM-P and P-FM boundaries are in close agreement with the Monte-Carlo simulations of Fig.~\ref{fig.raw_prob}. The asymptotic behavior in Fig.~\ref{fig.Lbif} for the AFM and P states at $J/\epsilon = 0.5$ and $J/\epsilon = 1$ respectively, is  a consequence of $\epsilon_J$ approaching zero, destabilizing the stationary solution. We note that the ramp-time of the pump can influence the phase diagram. Fast ramp times soften the competitive advantage of a low spin-bifurcation threshold, resulting in a blurring of the phase-boundaries (see Sec.~\ref{sec.app2}). The depression at low $W$ in Fig.~\ref{fig.raw_prob}(c) corresponds to an area of instability outlined by the black dashed line calculated using linear stability analysis for fluctuations at $k=0$ (see Sec.~\ref{sec.app1}).
\begin{figure}[!t]
\centering
	\includegraphics[width=0.48\textwidth]{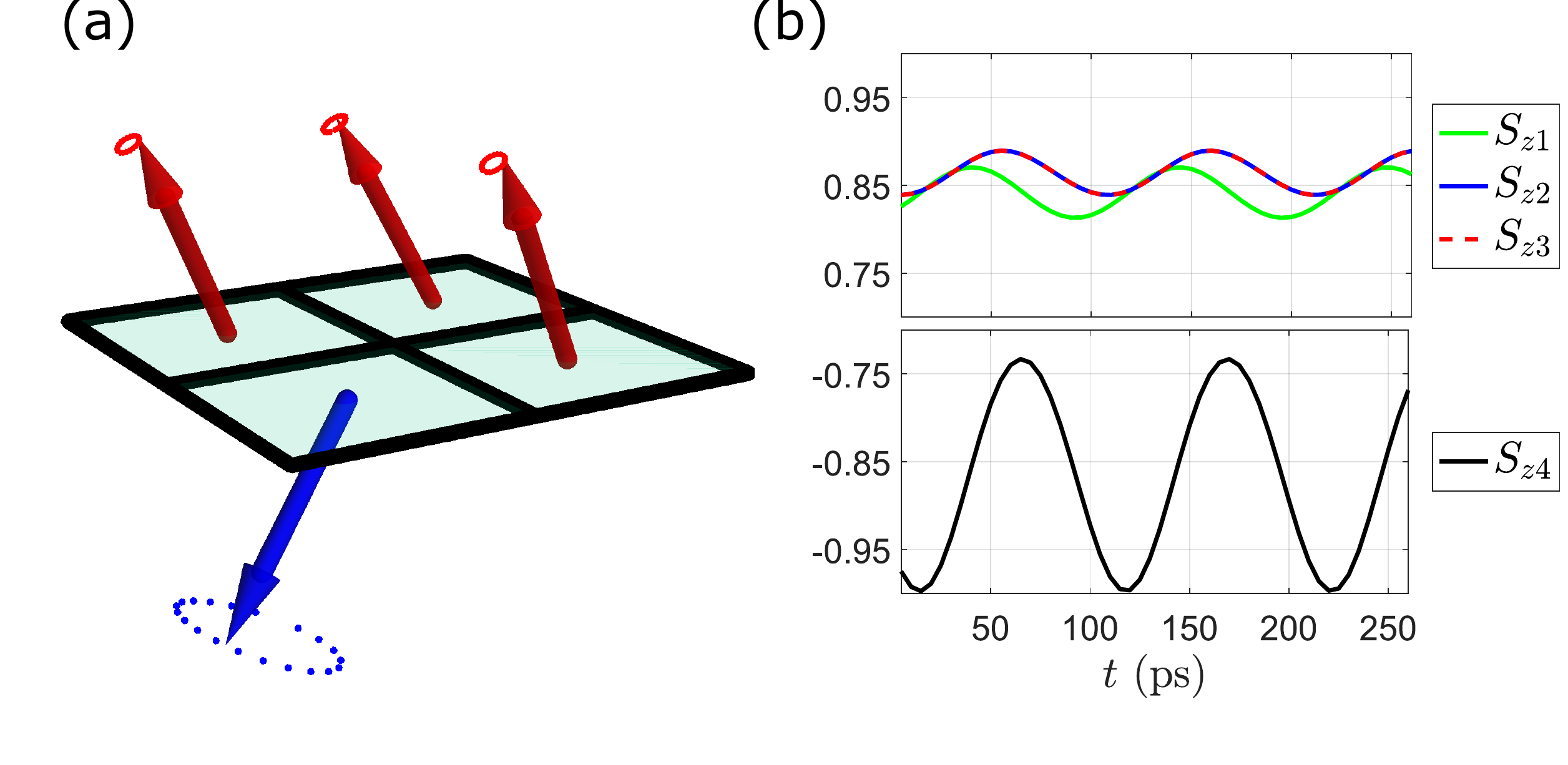}
\caption{(Color online) (a) Normalized pseudospins of the four condensates in the limit cycle solution sampled from the data in Fig.~\ref{fig.raw_prob}(d). Average spin has converged but multiple energies cause the pseudospins to precess. Plotted trajectories are derived from the right panel and have not been scaled. (b) Time evolution of this state over 260 ps showing the steady oscillation of the circular polarization component of the condensates.}
\label{fig.para2}
\end{figure}

We note that the MC iterations do not always result in a stationary AFM, P, or FM steady-state described above. Nonstationary symmetry-breaking solutions can arise close to stability boundaries due to the finite ramp rate of the pump. The analysis of highly nontrivial evolutions of the system in this case is beyond the scope of this work.

In addition to the stationary states, an oscillating limit cycle solution composed of three spins against one opposite spin is often observed for low $J$ and high $W$, as shown in Fig.~\ref{fig.raw_prob}(d) and Fig.~\ref{fig.para2}(a). A time-trace of the $S_z$ spin component of this state is plotted in Fig.~\ref{fig.para2}(b). Though the average spin on each site has converged, the spin precession indicates a superposition of states that are phase locked. Interestingly, the energy of the spinor-components of the limit cycle state correspond to that of P state, $\omega_J = -J$ and $\epsilon_J = \epsilon - J$, except for the minority spin population in the opposing condensate (e.g., $\Psi_{+4}$ polaritons from Fig.~\ref{fig.para2}) which also populates a separate peak in energy. Thus the limit cycle solution can be characterized as a `frustrated P-state', described by multiple energies $\omega_J$ and splittings $\epsilon_J$, and resulting in an oscillating spinor and frequency comb emission, similar to discussed in Ref.~\cite{rayanov_frequency_2015}. In larger chains, the limit cycle states can appear as a result of inhomogeneity in the chain couplings.

In conclusion, we have solved analytically and investigated numerically solutions in an infinite chain of coupled driven-dissipative spinor polariton condensates. A mixture of intra-spin coupling and nearest-neighbor inter-coupling allows not only controllable formation of antiferromagnetic states or ferromagnetic states, but also shows solutions with mixed antiferromagnetic and ferromagnetic bonding. We find that minimum bifurcation threshold determines the spin order in the chain. The one-to-one correspondence between the spin and the phase-slips of the lowest threshold states makes this system binary and opens the possibility of mapping it to binary models such as the 1D Ising Hamiltonian where the minimization of loss (bifurcation threshold) replaces minimization of energy. Our work is an important step towards understanding and controlling spin order in open-dissipative nonlinear spin lattices.

\emph{Acknowledgements.}---This work was supported by the Research Fund of the University of Iceland, The Icelandic Research Fund, Grant No.\ 163082-051, grant EPSRC EP/L027151/1, the Mexican Conacyt Grant No.\ 251808, and the Singaporean MOE grants 2015-T2-1-055 and 2016-T1-1-084. I.A.S. acknowledges support from a mega-grant \textnumero \ 14.Y26.31.0015 and
GOSZADANIE \textnumero \ 3.2614.2017/\fontencoding{T2A}\selectfont
ПЧ
\fontencoding{T1}\selectfont
 of the Ministry of Education and Science of
Russian Federation

\appendix
\begin{widetext}
\section{Complete solutions of 4 condensate chain system} \label{sec.app0}
It has been established that stationary chain solutions of either FM or AFM ordering results in a modified single-condensate dynamical equation (Eq.~\ref{eq:dotpsi2}) with only shifted parameters according to Eqs.~\ref{eq.epsomAFM}-\ref{eq.epsomFM}. Stationary chain solutions of mixed FM and AFM bonding (P solution) follow the same procedure but with only $\pi$ or zero phase slips possible between condensates. Focusing on a chain of 4 condensates (smallest cell to encompass all spin orderings periodically) we find 10 distinct solutions which are summarized in Fig.~\ref{fig.scheme3}. As the number of condensates in the chain increases, more solutions of FM or AFM ordering become available but the number of P solutions remains fixed. It's worth mentioning that panel (c) and (f) are special cases where the phases between neighboring condensates result in a cancellation such that $\omega_J = 0$ and $\epsilon_J = \epsilon$ in Eqs.~\ref{eq.epsJ}-\ref{eq.omJ}.
\begin{figure*}[!h]
  \centering
		\includegraphics[width=0.8\textwidth]{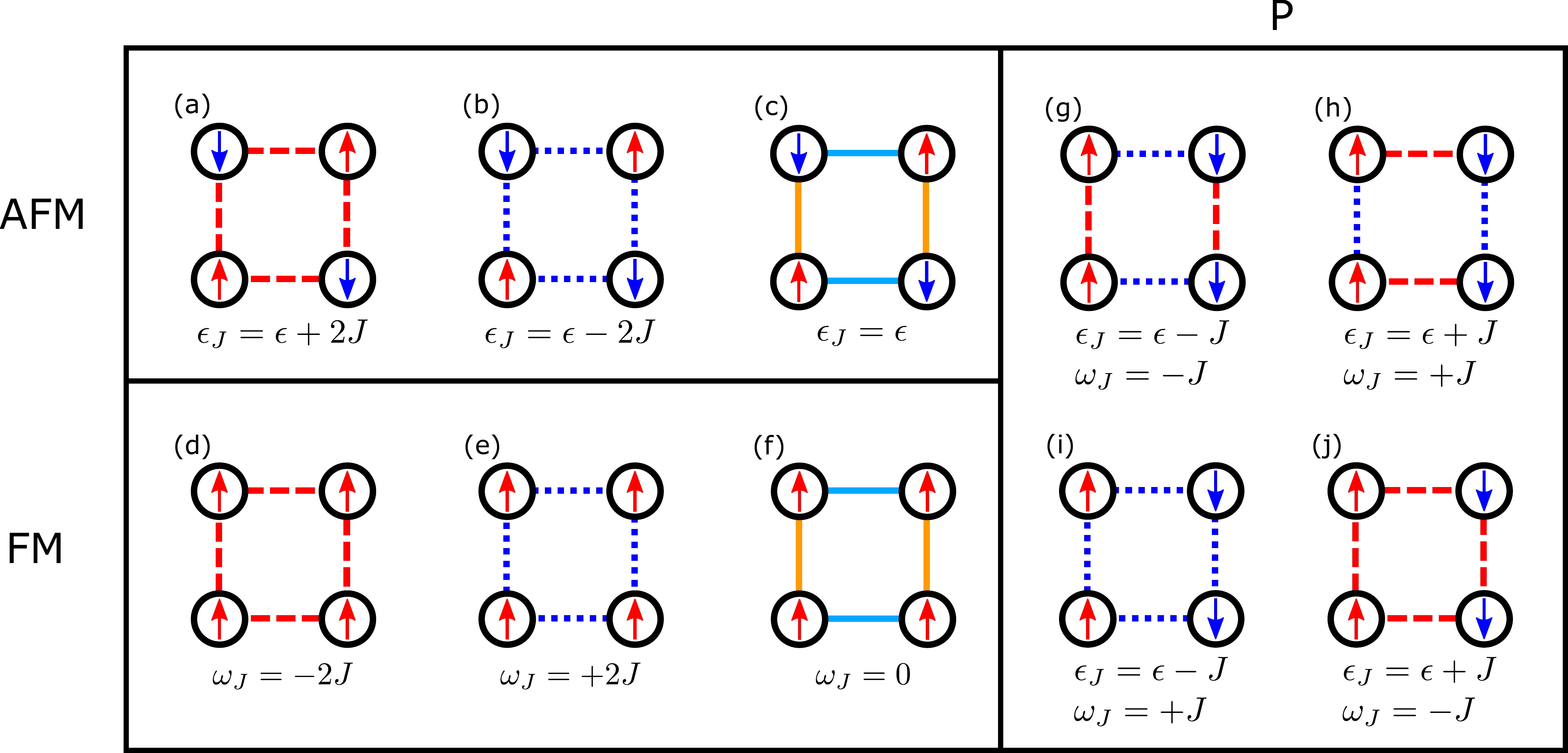}
	\caption{(Color online) Schematic showing 10 solutions of a 4 condensate chain system. Red dashed lines depict in-phase condensates ($\varphi_i = 0)$ and blue dotted lines anti-phase condensates ($\varphi = \pm\pi$). Orange and cyan whole lines in panels (c) and (f) correspond to phases causing a cancellation in the coupling, i.e., $e^{i \varphi_i} + e^{i \varphi_j} = 0$.}
\label{fig.scheme3}
\end{figure*}

\section{Linear stability analysis} \label{sec.app1}
In this section we formulate the linear stability analysis for a periodic solution for an infinite
chain of condensates. This solution is constructed by periodic repetition of a particular solution
for the closed ring of four condensate, and it has, in general, the period $4a$, where $a$ is the
nearest-neighbor distance. The perturbed solution can be written as
$\psi_\pm^{(m)}+\delta\psi_\pm^{(m)}$, where $m$ is the number of the condensate. The unperturbed
solution is periodic, $\psi_\pm^{(m)}=\psi_\pm^{(m+4)}$, and the perturbation is chosen in the
form of plane wave ($\hbar=1$)
\begin{equation} \label{eq.fluc}
 \delta\psi_\pm^{(m)} = u_\pm^{(m)} e^{ikma+\lambda t} + v_\pm^{(m)*} e^{-ikma+\lambda^* t}.
\end{equation}
Here the complex amplitudes are also set to be periodic, $u_\pm^{(m)}=u_\pm^{(m+4)}$ and
$v_\pm^{(m)}=v_\pm^{(m+4)}$. There are 16 linearized equations for the amplitudes and 16 Lyapunov
exponents $\lambda(k)$. The solution is stable when all of them satisfy $\Re\{\lambda(k)\}\le0$.

The linearized equations for the vector
$\mathbf{U}=\{u_+^{(1)},v_+^{(1)},u_-^{(1)},v_-^{(1)},\dots,u_-^{(4)},v_-^{(4)}\}^\mathrm{T}$ can
be written in matrix form as $i\lambda\mathbf{U}=\mathbf{M}\cdot\mathbf{U}$, where the $16\times16$
matrix $\mathbf{M}$ can be presented in the $4\times4$ blocks form:
\begin{equation} \label{eq.stab_mat_final}
\mathbf{M} = \begin{pmatrix} \mathcal{M}^{(1)} & \mathcal{M}_J &  0_{4,4} &  \mathcal{M}_J  \\
\mathcal{M}_J  &  \mathcal{M}^{(2)} &   \mathcal{M}_J & 0_{4,4}   \\
 0_{4,4} & \mathcal{M}_J  & \mathcal{M}^{(3)} &  \mathcal{M}_J  \\
 \mathcal{M}_J & 0_{4,4} & \mathcal{M}_J & \mathcal{M}^{(4)}   \end{pmatrix}.
\end{equation}
Here, $\mathcal{M}^{(n)}$ is the matrix describing the fluctuations in the $n$-th condensate
within the elementary cell, $n=1,2,3,4$, $0_{4,4}$ is the $4\times4$ zero matrix, and
$\mathcal{M}_J$ is the same-spin coupling between nearest neighboring condensates.

For the matrices in \eqref{eq.stab_mat_final} we have
\begin{align}
\mathcal{M}^{(n)} = \mathcal{M}_E^{(n)} + \mathcal{M}_W^{(n)} + \mathcal{M}_\gamma + M_\epsilon + M_{\alpha_1}^{(n)} + M_{\alpha_2}^{(n)}.
\end{align}
The first matrix is defined by the energy of $n$-th condensate
\begin{align}\mathcal{M}_E^{(n)} =
\begin{pmatrix}
-E^{(n)} & 0  &   0 & 0 \\
0 & E^{(n)} & 0  &  0\\
 0 & 0  & -E^{(n)} &  0 \\
0  & 0 &  0 &  E^{(n)}
\end{pmatrix}.
\end{align}
The second matrix arises from the harvest and saturation rates of the condensate from the static
reservoir
\begin{equation}\small
 \mathcal{M}_{W}^{(n)} =  -\frac{i \eta }{4 } \begin{pmatrix}
2|\psi_+^{(n)}|^2 +|\psi_-^{(n)}|^2  &  (\psi_+^{(n)})^2 & \psi_+^{(n)}\psi_-^{(n)*} & \psi_+^{(n)}\psi_-^{(n)}  \\
(\psi_+^{(n)*})^2 &  2|\psi_+^{(n)}|^2+|\psi_-^{(n)}|^2  & \psi_+^{(n)*}\psi_-^{(n)*} & \psi_+^{(n)*}\psi_-^{(n)} \\
\psi_+^{(n)*}\psi_-^{(n)} & \psi_+^{(n)}\psi_-^{(n)} &  2|\psi_-^{(n)}|^2+ |\psi_+^{(n)}|^2 &  (\psi_-^{(n)})^2 \\
\psi_+^{(n)*}\psi_-^{(n)*} &  \psi_+^{(n)}\psi_-^{(n)*} &   (\psi_-^{(n)*})^2 &
2|\psi_-^{(n)}|^2 + |\psi_+^{(n)}|^2 \end{pmatrix} - \frac{i}{2}G \mathbb{I},
\end{equation}
where $G = \Gamma-W$ and $\mathbb{I}$ is the identity matrix. The third and the fourth matrices
arise from coupling between up and down components
\begin{alignat}{2}
\mathcal{M}_\gamma =  \frac{i  \gamma}{2}
\begin{pmatrix}
0 & 0 & -1  & 0
\\
0 & 0 & 0  & -1 \\
-1  & 0  & 0 &  0 \\
0  &  -1 &  0  &   0
\end{pmatrix},
& \qquad
   & \mathcal{M}_\epsilon =
\frac{ \epsilon}{2} \begin{pmatrix}
0 & 0  &   -1 & 0 \\
0 & 0 & 0  &    1\\
 -1 & 0  & 0 &  0 \\
0  &   1 &  0 &  0
\end{pmatrix}.
\end{alignat}
The fifth and the sixth matrices arise from the interactions: {\small
    \begin{align}
        \mathcal{M}_{\alpha_1}^{(n)} &= \frac{\alpha_1}{2} \begin{pmatrix} 2|\psi_+^{(n)}|^2  &  (\psi_+^{(n)})^2 & 0 & 0 \\
			  		- (\psi_+^{(n)*})^2 & - 2|\psi_+^{(n)}|^2  & 0 & 0 \\
					0 & 0 &  2|\psi_-^{(n)}|^2&  (\psi_-^{(n)})^2 \\
					0 &  0 & -  (\psi_-^{(n)*})^2 & -  2|\psi_-^{(n)}|^2 \end{pmatrix},
                    \nonumber \\
 \mathcal{M}_{\alpha_2}^{(n)} &= \frac{ \alpha_2}{2} \begin{pmatrix}
|\psi_-^{(n)}|^2  & 0 & \psi_+^{(n)}\psi_-^{(n)*} & \psi_+^{(n)}\psi_-^{(n)} \\
0 & - |\psi_-^{(n)}|^2  & -\psi_+^{(n)*}\psi_-^{(n)*} & -\psi_+^{(n)*}\psi_-^{(n)} \\
\psi_+^{(n)*}\psi_-^{(n)} & \psi_+^{(n)}\psi_-^{(n)} &  |\psi_+^{(n)}|^2&  0 \\
-\psi_+^{(n)*}\psi_-^{(n)*} &  -\psi_+^{(n)}\psi_-^{(n)*} &0 & -
|\psi_+^{(n)}|^2 \end{pmatrix}.
\end{align}
} The final matrix describes the same-spin coupling between the nearest-neighboring condensates
\begin{align} \label{eq.matJ}
\mathcal{M}_J = \frac{\cos{(ka)}}{2} \begin{pmatrix}
-J& 0  &   0 & 0 \\
0 & J^* & 0  &   0\\
0  & 0  & -J &  0 \\
0  &  0 &  0 &  J^*
\end{pmatrix}.
\end{align}

\section{Non-adiabatic Monte-Carlo for undamped 4 condensate chain} \label{sec.app2}
 Here we give results analogous to Fig.~\ref{fig.raw_prob} but with damping absent ($\Im{(J)} = 0$) and instantaneous switching of the pump intensity at its mark value. Unlike Fig.~3 where states with the lowest bifurcation threshold were dominant, we uncover a more complex probability map in Fig.~\ref{fig.raw_prob2} through 30 MC trials bounded by their $k=0$ stability regions (black dashed lines) predicted by Eq.~\ref{eq.stab_mat_final}.
\begin{figure}[!t]
  \centering
		\includegraphics[width=0.8\textwidth]{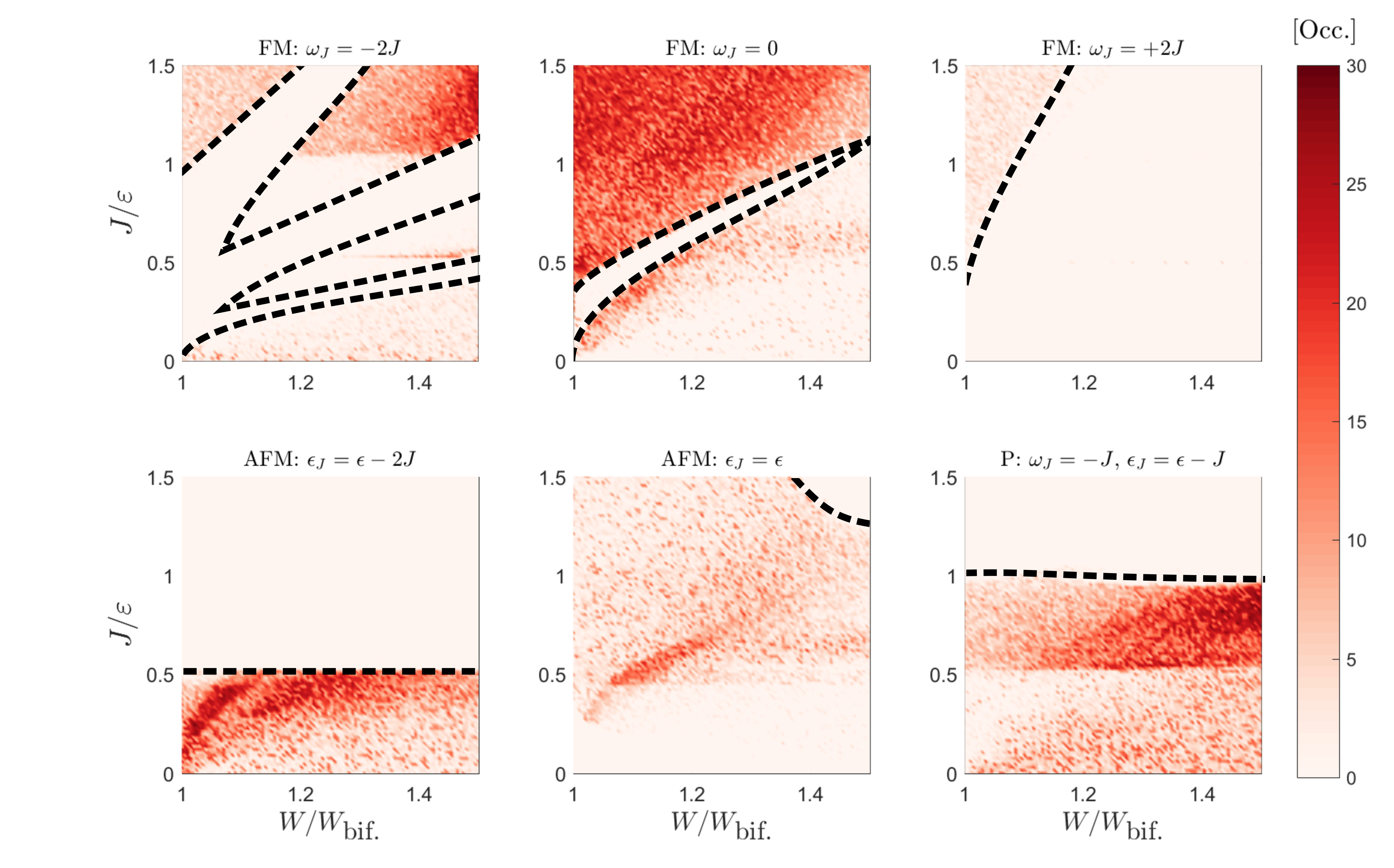}
	\caption{(Color online) Colormaps showing the likelihood of a spin state appearing through 30 MC iterations for each pixel in a $100 \times 100$ map using Eq.~\ref{eq.fullGP}. Different from Fig.~\ref{fig.raw_prob} we get a noticeable population in three more states (FM: $\omega_J = 0$. AFM: $\epsilon_J = \epsilon$. FM: $\omega_J = +2J$). Black dashed lines are predicted stability boundaries calculated using Eq.~\ref{eq.stab_mat_final} for $k=0$.}
\label{fig.raw_prob2}
\end{figure}
 
Fig.~\ref{fig.raw_prob2} shows 77\% of the data divided between 6 solutions of the 4 condensate chain. The remaining 4 solutions from Fig.~\ref{fig.scheme3} are not observed since they are unstable over the entire $J-W$ map. Another 11\% of the data (not shown here) ended in oscillating limit cycle states discussed in Fig.~\ref{fig.para2}. The remaining 12\% were categorized as nonstationary/chaotic.

The complex features of the probability maps in Fig.~\ref{fig.raw_prob2} as opposed to the more simplistic ones in Fig.~\ref{fig.raw_prob} highlight the important role of damping in the system and adiabatic switching of the pump intensity. Intuitively, the complicated features in Fig.~\ref{fig.raw_prob2} arise from the order parameter overshooting many possible stable minima in the phase space of the system. It then becomes a matter of the nearest and strongest attractor to stabilize the solution.

\section{Numerical methods} \label{sec.app3}
A QR algorithm is implemented to solve the eigenvalue problem of Eq.~\ref{eq.stab_mat_final}. Eq.~\ref{eq.fullGP} is solved using a variable-order Adams-Bashforth-Moulton predictor-corrector method. The parameters used for 0D simulations were: 
$\eta=0.02$  ps$^{-1}$; 
$\Gamma=0.1$ ps$^{-1}$; 
$\epsilon=0.04$ ps$^{-1}$; 
$\gamma=0.2\epsilon$;
$\alpha_1= 0.01$ ps$^{-1}$; 
$\alpha_2=-0.5\alpha_1$.

The parameters used for 2D simulations of Eq.~\ref{eq.fullGP2D} were: 
$m^* = 5 \times 10^{-5} m_0$;
$\eta=0.01$  ps$^{-1}$; 
$\Gamma=0.2$ ps$^{-1}$; 
$\epsilon=0.015$ ps$^{-1}$; 
$\gamma=0.2\epsilon$;
$\alpha_1= 0.003$ ps$^{-1}$; 
$\alpha_2=-0.5\alpha_1$;
$d = 12$ $\mu$m;
$\sigma = 10.3$ $\mu$m.
Where $m_0$ is the free electron rest mass.

\section{Derivation of the coupling contribution in chain systems} \label{sec.app4}
Consider the stationary condensate chain composed entirely of either FM- or AFM bonds. According to Eqs.~\ref{eq.epsJ}-\ref{eq.omJ}, each condensate with two nearest neighbors is presented with a term
\begin{equation} \label{eq1}
\omega_J = -J (e^{i \varphi_i} + e^{i\varphi_j}),
\end{equation}
for two FM bonds or
\begin{equation} \label{eq2}
\epsilon_J = \epsilon + J (e^{i \varphi_i} + e^{i\varphi_j}),
\end{equation}
for two AFM bonds. Here $\varphi_{i,j}$ is the phase difference of moving from the condensate in question to its neighbor. This contribution can in general be a complex number appearing equally in each condensate. In this section we show that this number must stay real for a chain system. The lattice unit cell of the chain system is one condensate and one bond. Assuming that the chain closes on itself, the number of free variables ($\varphi_i$) is then equal to the number of independent equations. The following can then be generalized to any number of condensates in a chain. 

Let's now imagine 4 condensates locked in a chain. We can classify the phase jumps going clockwise as $\{\varphi_1, \varphi_2, \varphi_3, \varphi_4\}$ (see Fig.~\ref{fig}). It is obvious that $\omega_J$ and $\epsilon_J$ must be equal for all condensates in the chain in order to have a steady state. Thus the phase contribution $e^{i \varphi_i} + e^{i \varphi_j}$ must be the same for all condensates. This means that the 4 stationary condensates allow us to write,
\begin{equation}
e^{i \varphi_1} + e^{-i\varphi_4} = e^{i \varphi_2} + e^{-i\varphi_1} = e^{i \varphi_3} + e^{-i\varphi_2} =e^{i \varphi_4} + e^{-i\varphi_3}.
\end{equation}
\begin{figure}[!t]
\centering
\includegraphics[width = 0.75\linewidth]{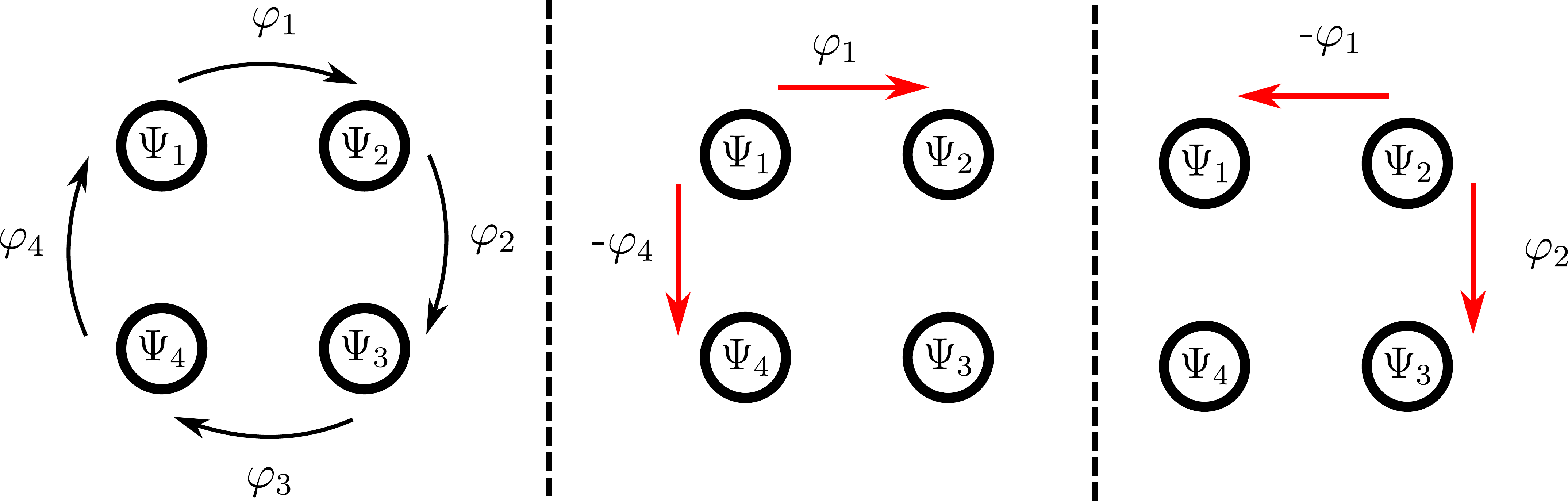}
\caption{Schematic of the 4 condensate chain. (Left) Phase jumps $\varphi_i$ take place moving from one condensate to the next. (Middle) $\Psi_1$ gets a contribution $e^{i\varphi_1} + e^{-i \varphi_4}$. (Right) $\Psi_2$ gets a contribution $e^{i \varphi_2} + e^{-i \varphi_1}$.}
\label{fig}
\end{figure}
Writing $e^{i \varphi_i} = a_i + i b_i$ where $a_i,b_i \in \mathbb{R}$ it's then easy to show that,
\begin{align}
& a_1 = a_3, \\
& a_2 = a_4, \\
& b_1 = b_2 = b_3 = b_4. 
\end{align}
The real part of the contribution $e^{i \varphi_i} + e^{i \varphi_j}$ is thus equal for each condensate but the imaginary part gets canceled. Consequently, from $|e^{i \varphi_i}|^2 = 1$ we come to the solution $a_1 = \mp a_2$, which can more clearly be written:
\begin{equation} \label{eq.sol0}
\cos{(\varphi_1)} = \mp \cos{(\varphi_2)}.
\end{equation}
The minus sign in Eq.~\ref{eq.sol0} corresponds to a cancellation in the coupling with no shift in $\omega_J$ or $\epsilon_J$ whereas the plus sign mandates the opposite. Applying the constraint $e^{i(\varphi_1 + \varphi_2 + \varphi_3 + \varphi_4)} = 1$ corresponding to a full cycle in our 4 condensate chain we come to the conclusion that the only possible values of coupling in the latter case are $\cos{(\varphi_1)} = \cos{(2 \pi m / N)}$ where $m = 0,1,2,\dots$ and $N$ is the number of condensates in the chain. As the number of condensates increases in the chain, more solutions become available.

The same procedure can be applied to a state where each condensate has one FM- and one AFM bond (P solutions). Then only $e^{i\varphi_i} = \pm1$ satisfies the chain.

\section{4 condensate chain with spatial degrees of freedom} \label{sec:app6}
The tight binding model (Eq.~\ref{eq.fullGP}) offers a simple solution to the stationary spin patterns in the condensate chain. We find that these exact solutions can also be produced with little difficulty accounting for the spatial degree of freedom. The complex Ginzburg-Landau equation can be written then~\cite{keeling_spontaneous_2008, wouters_excitations_2007, wouters_energy_2010}:
\begin{equation} \label{eq.fullGP2D}
		i\dot{\Psi}= \frac{1}{2} \left[ -i \left( g(S) + \gamma\sigma_x \right)  +(1-i\Lambda)\left( \bar{\alpha} S  + \alpha S_{z}\sigma_z   - \epsilon\sigma_x -\frac{\hbar \nabla^2}{m^*}+ g_P P(\mathbf{r}) \right) \right] \Psi.
\end{equation}
Here, $m^*$ is the effective mass of the polaritons. The exciton reservoir is taken to be completely static and the induced repulsive potential is then given by an effective interaction constant $g_P$. The remainder of the parameters serve the same purpose here as in the tight binding model with $W = P(\mathbf{r})$. We note that modeling the system by coupling an exciton reservoir rate equation to the order parameter only requires rescaling of the parameters and does not critically affect the observed solutions in Eq.~\ref{eq.fullGP2D} when the decay rate of the reservoir is taken to be large compared to the polariton lifetime~\cite{Smirnov_dark_solitons}. 

The pump $P(\mathbf{r})$ is a $3 \times 3$ square-arrangement of Gaussians separated by the lateral distance $d$ and with a FWHM $\sigma$ which then form four potential minimum in a $2\times2$ arrangement where the polaritons condense. In Fig.~\ref{fig.2D solutions} we show three solutions of different spin order in a closed 4 condensate chain. The AFM, P, and FM solutions possess phase slips corresponding to the lowest bifurcation threshold solutions from Fig.~\ref{fig.scheme3}(b,d,g), and Fig.~\ref{fig.raw_prob}(a-c). Each solution can be achieved by either tuning the strength of polariton interaction with the pump $g_P$, or by increasing the strength of the center Gaussian pump spot causing an increased barrier between the condensates which effectively changes the coupling strength $J$. 
\begin{figure}[!h]
  \centering
		\includegraphics[width=0.8\textwidth]{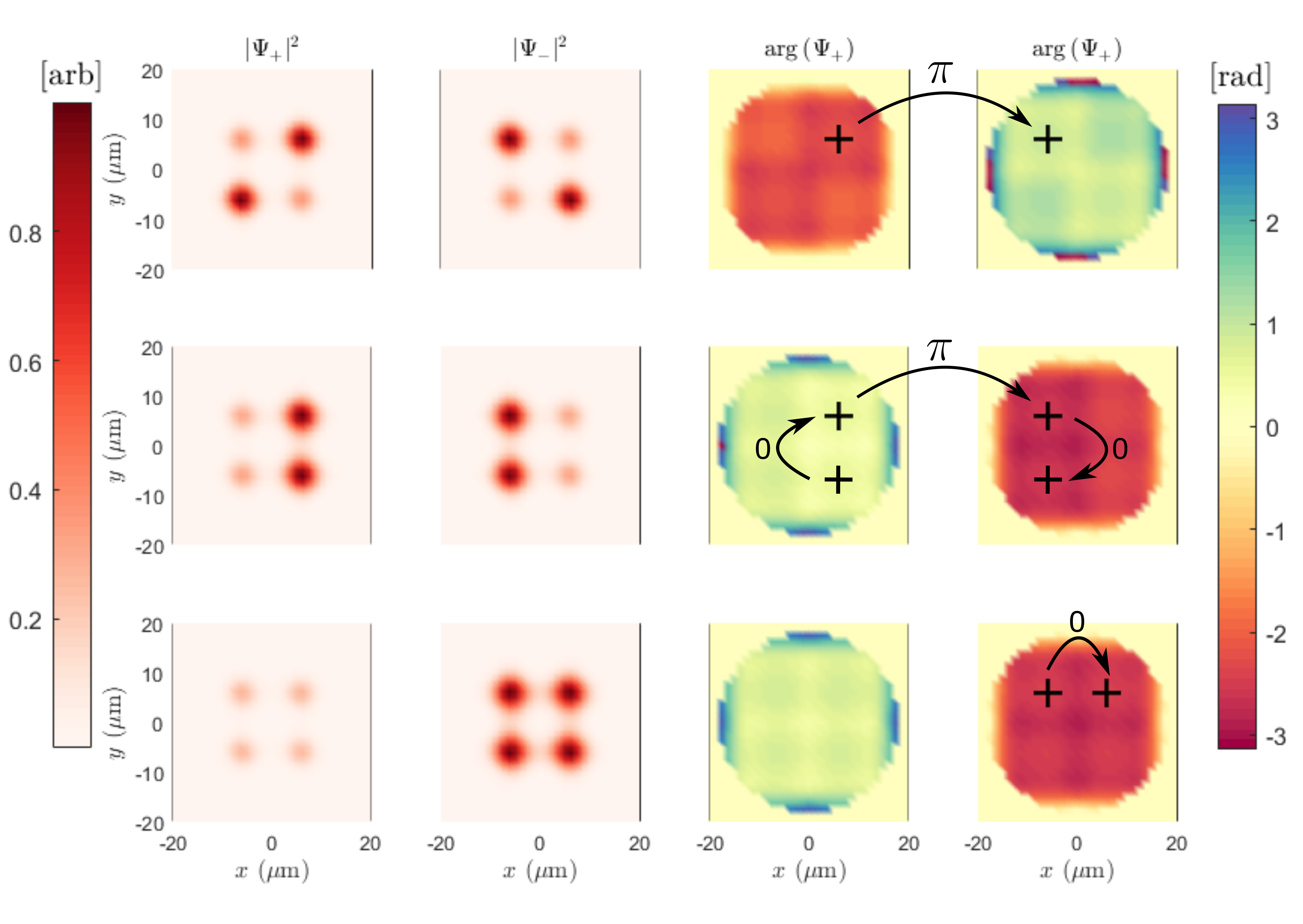}
	\caption{(Color online) Density and phase maps of the AFM, P, and FM spin states from Eq.~\ref{eq.fullGP2D}. Evaluating the phase difference between the black crosses in each solution confirms that AFM bonds favor $\pi$ phase difference and FM bonds 0 phase difference.}
\label{fig.2D solutions}
\end{figure}

\end{widetext}

\bibliography{Bibliography}

\end{document}